\documentclass[twocolumn,aps,prl,showpacs,superscriptaddress]{revtex4}

\usepackage{graphicx}

\begin{document}
\title{Experimental quantum ``Guess my Number'' protocol using multiphoton entanglement}
\author{Jun Zhang}
\email{zhangjun@mail.ustc.edu.cn}
\affiliation{Hefei National Laboratory for Physical Sciences at Microscale and Department
of Modern Physics, University of Science and Technology of China, Hefei, Anhui
230026, China}
\author{Xiao-Hui Bao}
\affiliation{Hefei National Laboratory for Physical Sciences at Microscale and Department
of Modern Physics, University of Science and Technology of China, Hefei, Anhui
230026, China}
\author{Teng-Yun Chen}
\affiliation{Hefei National Laboratory for Physical Sciences at
Microscale and Department of Modern Physics, University of Science
and Technology of China, Hefei, Anhui 230026, China}
\author{Tao Yang}
\email{yangtao@ustc.edu.cn}
\affiliation{Hefei National Laboratory for Physical Sciences at Microscale and Department
of Modern Physics, University of Science and Technology of China, Hefei, Anhui
230026, China}
\author{Ad\'{a}n Cabello}
\email{adan@us.es}
\affiliation{Departamento de F\'{\i}sica Aplicada II, Universidad de Sevilla, 41012 Sevilla, Spain}
\author{Jian-Wei Pan}
\email{jian-wei.pan@physi.uni-heidelberg.de}
\affiliation{Hefei National Laboratory for Physical Sciences at Microscale and Department
of Modern Physics, University of Science and Technology of China, Hefei, Anhui
230026, China}
\affiliation{Physikalisches Institut, Universit\"at Heidelberg, Philosophenweg 12, 69120 Heidelberg, Germany}
\date{\today}


\begin{abstract}
We present an experimental demonstration of a modified version of
the entanglement-assisted ``Guess my Number'' protocol for the
reduction of communication complexity among three separated parties.
The results of experimental measurements imply that the separated
parties can compute a function of distributed inputs by exchanging
less classical information than by using any classical strategy. And
the results also demonstrate the advantages of entanglement-enhanced
communication, which is very close to quantum communication. The
advantages are based on the properties of
Greenberger-Horne-Zeilinger states.
\end{abstract}


\maketitle


One of the most challenging applications of quantum mechanics for
information processing is the reduction of the amount of
communication needed to compute a function of a number of inputs
distributed between distant
parties~\cite{CB97,BCW98,BvHT99,Raz99,Sv00,XHZLG01,TSBBZW05,Y79}.
Compared to the classical scenario, the quantum scenario with the
assistance of quantum entanglement has significant advantages; i.e.,
less classical information is needed from one place to another than
the classically required amount of communication, or the quantum
scenario can reduce the communication complexity. A particularly
attractive and stimulating way of showing the quantum advantages in
a multiparty scenario was proposed by Steane and van Dam as a method
for always winning the television contest ``Guess my Number''
(GMN)~\cite{Sv00}. Steane and van Dam stressed that ``A laboratory
demonstration of entanglement-enhanced communication would be
\ldots a landmark in quantum physics and quantum information
science''~\cite{Sv00}.


Previous experiments have demonstrated the reduction of two-party
communication complexity by using two-photon nonmaximally
entangled states~\cite{XHZLG01} and the reduction of $N$-party
communication complexity using a single
qubit~\cite{TSBBZW05}. However, the GMN game can provide a more direct and
effective demonstration of the reduction of multiparty
communication complexity. Though the high detection efficiencies
required for the original GMN game~\cite{Sv00} have
prevented further progress, recently the study of
modified versions of the GMN game requiring lower detection
efficiencies has prompted~\cite{CL05}.

In this paper we demonstrate the reduction of multiparty
communication complexity by using genuine multiqubit
entanglement. In the experiment, a three-party quantum GMN
protocol using multiphoton entanglement is demonstrated to
implement a significant reduction of the communication complexity
between three separated parties. The protocol is based on a
further modification of the GMN game which preserves all the
essential features of the original game. The quantum advantages
are based on the properties of a polarization-entangled
Greenberger-Horne-Zeilinger (GHZ) state~\cite{GHZ89,GHSZ90}.


The modified GMN game is as follows: a team of three contestants,
$Alice$, $Bob$, and $Charlie$, plays against a TV program's host.
Before the game starts, the contestants decide on a common
strategy. Each contestant is then isolated in a separate booth.
They can take anything they want with them into the booths, except
clocks or devices which would allow any two of them to share a
temporal reference~\cite{Sv00}. At random moments, the host gives
the contestant $j$ ($j=A, B, C$) a randomly chosen number $n_j=0$, $1/2$, $1$,
or $3/2$ of apples, so that the sum of all three, $n=n_A+n_B+n_C$,
is an integer number. The team's task is to ascertain whether $n$
is even or odd. Each contestant gives the host a bit value $b_j$
and the team's answer to the question ``what is the parity of
$n$?'' is the sum modulo two of these three bits,
$b=(b_A+b_B+b_C)~mod~2$. ``$b=0$'' means that they think $n$ is an
even number, and ``1'' means that they think $n$ is an odd number.
The contestants are permitted to refuse to answer (i.e., not to
give the host a bit) in any round. The host is permitted to give
apples to some of the contestants only. Valid rounds are those in
which the host has given apples to all three contestants and all
of them have given a bit to the host. The host must guarantee that
valid rounds are equally distributed among the $32$~possible
variations of apples.

Considering the above rules, since the contestants can refuse to
give bits to the host, the first possible classical strategy for
them is to use fixed local instructions like ``give the bit
$b_j=0$ upon receiving $n_j=0$ apples, give $b_j=1$ upon receiving
$n_j=1$ apples, and give nothing in other cases.'' Of course, the
host can recognize this strategy easily. If the host does not
prevent their strategy, as a result the team definitely wins the
game with a probability of $100\%$. However, in order to guarantee
that valid rounds are equally distributed among the possible
variations of apples, the host will insist on those variations
that the contestants are refusing to give bits, so that the
contestants cannot take advantage of this possibility.


Another possible classical strategy is to use a secret sequence of
local instructions like ``give $b_j=0$ upon receiving $n_j=0$
apples, give $b_j=1$ upon receiving $n_j=1$ apples, and give
nothing in other cases, in rounds numbered two, seven and so on;
give \ldots in rounds numbered one, six and so on; etc.'' In
other words, the team could make a common table before the game,
which displays the corresponding instructions for every round, by
randomly selecting only two or three of the four numbers of apples
to give bits so that every possible variation appears in the table
with the same frequency. This strategy allows the team to win
every round if all the contestants agree on which round they are
partaking. However, the contestants cannot take advantage of this
strategy because none of them knows which valid round he (she)
currently finds himself (herself) in, since the host is permitted
to give apples to some of the contestants only, and also because
the contestants do not share temporal references, since clocks and
timing devices are forbidden.


The best classical strategies (i.e., those allowed by classical
physics) are previously decided local instructions like ``give
$b_j=1$ on receiving $n_j=0$ or $1/2$ apples, and give $b_j=0$ on
receiving $n_j=1$ or $3/2$ apples.'' A careful examination
reveals that this strategy, which allows the team to win the game with
probability of $3/4$, is indeed optimal.

Oppositely, the contestants can always win the game by using the
following entanglement-assisted protocol.

(1) Each contestant receives a photon belonging to a three-photon
entanglement system initially prepared in the GHZ states
\begin{equation}
|{\rm GHZ}\rangle = {1 \over \sqrt{2}} (|HHH\rangle+|VVV\rangle),
\label{GHZ03}
\end{equation}
where $|H\rangle$ and $|V\rangle$ represent horizontal and
vertical polarization, respectively.

(2) Each contestant $j$ applies to his photon the rotation
\begin{equation}
R(n_j)=|H\rangle\langle H|+e^{i n_j \pi}|V\rangle\langle V|,
\label{rotj}
\end{equation}
where $n_j$ is his number of apples.

(3) Each contestant then measures the polarization of his photon
in the $\{|+\rangle, |-\rangle\}$ basis, where $|+\rangle = {1 /
\sqrt{2}} (|H\rangle+|V\rangle)$ and $|-\rangle = {1 / \sqrt{2}}
(|H\rangle-|V\rangle)$.

(4) Sometimes, due to the inefficiency of the detectors,
contestant $j$ does not detect his photon. In these cases, he will
not give $b_j$ to the host. Note that the inefficiencies keep the
contestants from using the detections as a method to have common
references in time. When all contestants give the host a bit,
their sum modulo 2 is the correct answer, due to the following
property of state~(\ref{GHZ03}): for any $n_A+n_B+n_C$ integer
(where $n_j=0$, $1/2$, $1$, or $3/2$), the result of applying a
rotation to each photon is
\begin{eqnarray}
\lefteqn{R(n_A) \otimes R(n_B) \otimes R(n_C)
|{\rm GHZ}\rangle} \nonumber \\
& & = \left\{\begin{array}{ll}
|{\rm GHZ}\rangle & \mbox{if $n_A+n_B+n_C$ is even} \\
|{\rm GHZ}^\perp\rangle & \mbox{if $n_A+n_B+n_C$ is odd},
\end{array} \right.
\label{rule}
\end{eqnarray}
where $|{\rm GHZ}\rangle$ and $|{\rm GHZ}^\perp\rangle$ can be
reliably distinguished by local measurements in the $\{|+\rangle,
|-\rangle\}$ basis. This can be checked by rewriting the states in
that basis:
\begin{equation}
|{\rm GHZ}\rangle = {1 \over 2}(|+++\rangle+|+--\rangle+|-+-\rangle+|--+\rangle),
\end{equation}
\begin{equation}
|{\rm GHZ}^\perp\rangle = {1 \over 2}(|-++\rangle+|+-+\rangle+|++-\rangle+|---\rangle).
\end{equation}


Therefore, the contestants can always win the GMN game with the
assistance of quantum entanglement, while they can only win the
game with a probability (of $3/4$) without quantum entanglement.
Further analysis indicates that simulating the quantum advantage
would require two bits of communication between the contestants.
Therefore, the GMN game demonstrates that quantum entanglement can
be used to reduce the communication complexity.



\begin{figure}
\centerline{\includegraphics[scale=0.66]{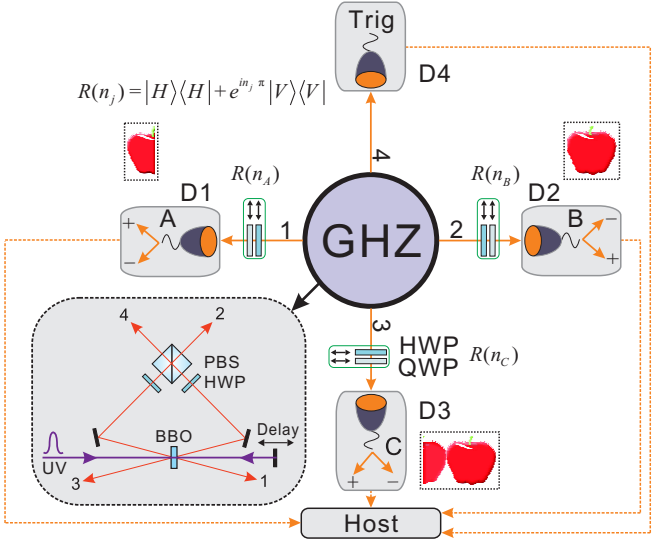}}
\caption{\label{Fig1} Experimental setup for the three-party GMN
game using three-photon entanglement. Two pairs of
polarization-entangled photons are produced by spontaneous
parametric down-conversion. A mode-locked Ti:sapphire femtosecond
laser, with pulse duration of 200\,fs and repetition rate of
76\,MHz at 788\,nm central wavelength, generates ultraviolet
(UV) pulses with 394\,nm wavelength, with an average pump power
of about 550\,mW, after a frequency-doubled process in a
$LiB_{3}O_{5}$ crystal. The UV pulses pass through a 2\,mm
thick BBO crystal twice to generate approximately 18 000 forward pairs
and 14 000 backward pairs of entangled photons with full width at half maximum (FWHW)=2.8\,nm
interference filters at 788\,nm.
One photon in each pair is overlapped
at the PBS temporally and spatially.
Three photons of modes 1, 2, and 3 are distributed to the
three separated parties $A$, $B$ and $C$ respectively,
while the photon polarization of mode 4 is fixed at $45^{\rm o}$
as a trigger for the four-fold coincidence. The
three parties use $R(n_{A})$, $R(n_{B})$ and $R(n_{C})$,
which are implemented using the combination of half-wave plate (HWP)
and quarter-wave plate (QWP),
to accomplish the rotations required for the quantum GMN protocol respectively.
Finally, three polarizers (POL) are used to implement the
projective measurements of the linear polarization of photons.}
\end{figure}

We have implemented the three-party quantum GMN protocol
using three-photon entanglement. The experimental setup is
shown in Fig.~\ref{Fig1}.

The first step is preparing three-photon GHZ states using
the techniques similar to those in previous
experiments~\cite{KMWZSS95,PBHZ98,PDGWZ01,ZYCZZP03}.
Ultraviolet pump pulses pass through a $\beta$-barium
borate (BBO) crystal twice to produce two pairs of
polarization-entangled photons. The states of both pairs after
passing through the two additional HWP's in modes $2$ and $4$ are
\begin{equation}
|\psi\rangle_{12}=|\psi\rangle_{34} = (1 / \sqrt{2})(|H H\rangle+|V
V\rangle).
\label{2state}
\end{equation}
By adjusting the position of a delay mirror, the two photons in modes
$2$ and $4$ simultaneously arrive at the polarizing beam splitter
(PBS), which transmits horizontally polarized photons while
reflects vertically polarized photons. The PBS is used as a
parity check: detectors D2 and D4 fire only when the inputs of the
PBS are both $H$ photons or both $V$ photons. Then we can produce the
following GHZ state
\begin{equation}
|\Psi\rangle_{1234}=(1 / \sqrt{2})(|HHHH\rangle+|VVVV\rangle).
\label{GHZ}
\end{equation}

To confirm that the state is in a coherent superposition,
we measure the coincident count rates of
$++++$ and $+++-$ as a function of the delay
mirror's position to observe the interference. After optimal
fitting we find the maximum visibility at zero delay is
approximately $86\%$, which indicates that the four photons are
indeed in a coherent superposition. When the photon of mode 4 is
projected to $45^{\rm o}$,
the state of remaining three photons will be a three-photon
GHZ state~\cite{PBDWZ00}
\begin{equation}
|\Psi\rangle_{123}=(1 / \sqrt 2)(|HHH\rangle+|VVV\rangle).\\
\label{GHZ3}
\end{equation}
Then we use this three-photon
GHZ state to demonstrate the three-party quantum GMN game.

The next step is that the host starts the game and distributes the apples.
During the game, at random moments, the host gives contestant $j$ ($j=A, B, C$)
a number $n_{j}$ of apples. In the experiment,
the three photons of
modes 1, 2, and 3 are distributed to the three contestants and
the contestant $j$ then performs the  rotation $R(n_j)$ according to the corresponding
number of apples, which is implemented using
a suitably chosen combination of a HWP and a QWP as follows:
\begin{eqnarray}
\left\{\begin{array}{ll}
R(0)=|H\rangle\langle H|+|V\rangle\langle V| & \mbox{nothing} \\
R(1/2)=|H\rangle\langle H|+i|V\rangle\langle V| & \mbox{a QWP at $0^{\rm o}$}\\
R(1)=|H\rangle\langle H|-|V\rangle\langle V| & \mbox{a HWP at $0^{\rm o}$} \\
R(3/2)=|H\rangle\langle H|-i|V\rangle\langle V| & \mbox{HWP+QWP at $0^{\rm o}$,}
\end{array} \right.
\label{Rot}
\end{eqnarray}
where half an apple is equivalent to a QWP at $0^{\rm o}$, while one apple is
equivalent to a HWP at $0^{\rm o}$. When a fourfold coincidence event is detected
it implies that all the three contestants have received the apples and given a bit to
the host, respectively, which is a valid round in the game.


\begin{figure}
\centerline{\includegraphics[scale=0.5]{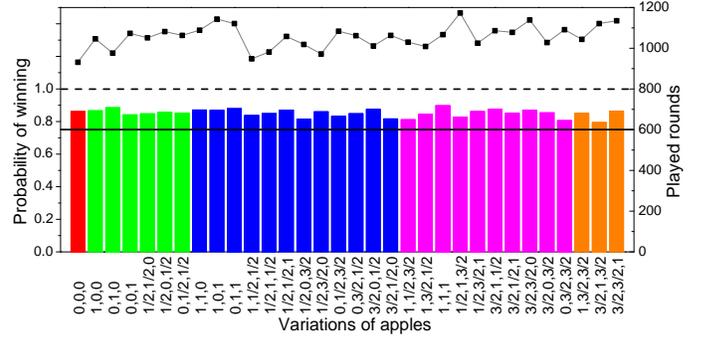}}
\caption{\label{Fig2} Measurement results of the quantum GMN
experiment for the $32$~possible variations of apples. The square
dots (corresponding to the right y axis) represent the number of
played rounds. Each variation has been played with approximately
the same frequency, in order to guarantee the game's fairness.
The histogram (corresponding to the left y axis) shows the
experimental probabilities of winning.
The dashed line (1.0) and the solid line (0.75)
represent the theoretical maximum quantum and classical
probabilities of winning the GMN game, respectively. The
experimental probabilities of winning the game are significantly higher than the
maximum classical probabilities.}
\end{figure}


During the game the $32$~possible variations of apples
should be the same frequency. In the experiment we produced the
$32$~cases through the different combinations of $R(n_j)$.
Each case was repeated for enough rounds
to reduce the statistical fluctuations and the total round numbers
of each case were almost identical. In each case, we
measured 8~kinds of coincidences (i.e., $+++$, $++-$, $+-+$,
$+--$, $-++$, $-+-$, $--+$, $---$) and the measurement time of
each coincidence was 30 min.

We recorded the $8$ numbers of counts $N_{+++}$,
\ldots, $N_{---}$ during $4$ h. According to Eq.~(\ref{rule}),
the number of rounds in which the players answer ``even'' is
\begin{equation}
N_{\rm even}=N_{+++}+N_{+--}+N_{-+-}+N_{--+}
\label{even}
\end{equation}
and the number of rounds in which the players answer ``odd'' is
\begin{equation}
N_{\rm odd}=N_{-++}+N_{+-+}+N_{++-}+N_{---}.
\label{odd}
\end{equation}
Therefore, the experimental probability of winning the GMN game is
$N_{\rm even}/(N_{\rm even}+N_{\rm odd})$ when $n$ is even and
$N_{\rm odd}/(N_{\rm even}+N_{\rm odd})$ when $n$ is odd. The
experimental results are shown in Fig.~\ref{Fig2}. In all of the
32~variations of apples the experimental probability of winning
is higher than the best classical value of $0.75$.
The number of played rounds of each possible
is about $1000$. In each variation of apples, the probability
of winning is about $0.85$ and differs from the best classical value by
about $9$ standard deviations. On the other hand, the total number
of rounds in which the answer was correct is $28 768$,
while the number of rounds in which the answer was incorrect is
$5 032$. Therefore, the mean value of the experimental probability
of winning is $\overline{P_{Q}}=0.851 \pm 0.002$, which differs
from the classical result with more than $52$ standard deviations.
This clearly illustrates the advantage of the entangled-assisted
strategy. The imperfection of $\overline{P_{Q}}$ (the theoretical
quantum prediction is $P_{Q}=1$) is mainly due to the visibility limitation
of multiphoton entanglement and a slight drift of the interference
position in the experiment. During the experiment,
the laser system cannot be stabilized for enough long time.
Therefore the best interference position cannot always be fixed at the
same point for enough time.


In conclusion, we have performed an experimental
implementation of a quantum protocol which outperforms the best
classical strategy for wining a modified version of the GMN game
preserving all the essential features of the original one.
Our results demonstrate the advantages of entanglement-assisted
communication and confirm one of the most challenging predictions
of quantum information processing and quantum
computation~\cite{NC00,BEZ00,MPZ00}. The experimental triumph in
the GMN game shows that entanglement is a physical resource that
can be used to reduce the classical communication cost of some
distributed computations in a multiparty scenario. The
entanglement-assisted reduction of classical communication
complexity has a number of potential applications in computer
networks, very large scale integrated (VLSI) circuits, and data structures~\cite{KS97}, and
deserves further research.


This work was supported by the NNSF of China, the CAS, the PCSIRT,
and the National Fundamental Research Program (under Grant No. 2001CB309303).
It was also supported by the Marie Curie Excellence Grant Program of
the European Union, the Alexander von Humboldt Foundation, Spanish MCT Project
No. BFM2002-02815, and Junta de Andaluc\'{\i}a Project No. FQM-239.



\end{document}